\documentclass[aps,prl,superscriptaddress,amsmath,amssymb,amsfonts,twocolumn,showpacs]{revtex4-2}
\usepackage{amsmath,amssymb,amsfonts,graphicx}
\usepackage{times,mathpazo,charter}
\usepackage[english]{babel}
\usepackage[usenames,dvipsnames]{xcolor}
\usepackage{stmaryrd}
\newcommand\orcid[1]{\href{https://orcid.org/#1}{\includegraphics[width=10pt]{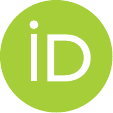}}}
\usepackage[colorlinks=true,citecolor=blue]{hyperref}
\graphicspath{{figures/}}

\usepackage[normalem]{ulem}

\newcommand{\fl}{f^{{\rm l}}}
\newcommand{\fr}{f^{{\rm r}}}

\newcommand{\al}{a^{{\rm l}}}
\newcommand{\ar}{a^{{\rm r}}}
\newcommand{\Fl}{F^{{\rm l}}}
\newcommand{\Fr}{F^{{\rm r}}}

\newcommand{\ceffl}{s^{\rm l}}
\newcommand{\ceffr}{s^{\rm r}}
\newcommand{\cl}{c^{{\rm l}}}
\newcommand{\crr}{c^{{\rm r}}}
\def\[{\begin{equation}}
\def\]{\end{equation}}

\def\sech{\mathop{\rm sech}\nolimits}
\def\sgn{\mathop{\rm sgn}\nolimits}
\def\bse{\begin{subequations}}
\def\ese{\end{subequations}}

\newcommand{\dd}{{\rm d}}

\begin{document}
\def\thetitle{Two-dimensional stationary soliton gas}
\title{\thetitle}

\author{Thibault Bonnemain\orcid{0000-0003-0969-2413}} 
\affiliation{Department of Mathematics, King's College, London, United Kingdom}

\author{Gino Biondini\orcid{0000-0003-3835-1343}}
\affiliation{Department of Mathematics and Department of Physics, State University of New York, Buffalo, NY, United States of America}

\author{Benjamin Doyon\orcid{0000-0002-5258-5544}}
\affiliation{Department of Mathematics, King's College, London, United Kingdom}

\author{Giacomo Roberti \orcid{0000-0001-8233-9531}}
\affiliation{Department of Mathematics, Physics and Electrical Engineering, Northumbria University, Newcastle upon Tyne, United Kingdom}

\author{Gennady A. El\orcid{0000-0003-1962-5388}} 
\affiliation{Department of Mathematics, Physics and Electrical Engineering, Northumbria University, Newcastle upon Tyne, United Kingdom}

\date{\small\today}

\begin{abstract}
We study two-dimensional stationary soliton gas in the framework of the time-independent reduction of the Kadomtsev-Petviashvili (KPII) equation, which coincides with the integrable two-way ``good'' Boussinesq equation in the $xy$-plane. This $(2+0)$D reduction enables the construction of the kinetic equation for the stationary gas of KP solitons by invoking recent results on $(1+1)$D bidirectional soliton gases and generalised hydrodynamics of the Boussinesq equation.  We then use the kinetic theory to analytically describe two basic types of 2D soliton gas interactions: (i) refraction of a line soliton by a stationary soliton gas, and (ii) oblique interference of two soliton gases.   We verify the analytical predictions by numerically implementing the corresponding  KPII soliton gases via exact $N$-soliton solutions with $N$-large and appropriately chosen random distributions for the soliton parameters. We also explicitly evaluate the long-distance correlations for the two-component interference configurations. The results can be applied to a variety of physical systems, from shallow water waves to Bose-Einstein condensates.
\end{abstract}

\maketitle

The concept of soliton gas (SG) as an infinite random ensemble of interacting solitons has recently emerged as a new and powerful tool for understanding random wave fields in the language of statistical physics and hydrodynamics. It was first introduced for a rarefied, or dilute, gas of Korteweg-de Vries (KdV) solitons  \cite{zakharov1971kinetic},
then generalized to dense gases and other integrable dynamical models \cite{el2003thermodynamic, el2005kinetic, el2020spectral, congy2021soliton}. 
SGs have been observed in various physical media, including water waves \cite{Costa:14, redor2019experimental, Suret2020Nonlinear, leduque2024space}, photorefractive crystals \cite{marcucci_topological_2019-1} and in superfluids \cite{mossman_observation_2023}.  

The theory of SGs relies on the integrability of the underlying nonlinear wave model and, in particular, on the tools of the inverse scattering transform (IST) that relate solitons to the discrete spectrum of the associated linear Lax operator \cite{ablowitz1981solitons, novikov1984theory}. 
A key object is the spectral kinetic equation for the density of states (DOS) ---  the joint distribution of solitons with respect to their spectral eigenvalues and positions. 

The kinetic equation for dense SGs can be interpreted 
%via a physically suggestive reasoning 
\cite{el2005kinetic} as a continuity equation for the DOS, where the flux density is determined by an integral equation of state that simply accounts for the average soliton velocity in the gas via a {\it collision rate ansatz}. The collision rate ansatz is obtained from two system-specific ingredients expressed in spectral terms: 
the free soliton velocity and the phase or position scattering shift. Soliton interactions are assumed to occur as sequences of two-body interactions, in agreement with two-particle factorisation of many-particle scattering. This is consistent with systematic spectral theory derivations for the KdV and focusing nonlinear Schr\"odinger (NLS)  SG \cite{el2003thermodynamic,el2020spectral}. The predictions of the kinetic SG theory for the focusing NLS equation were successfully verified in fibre optics and deep-water tank experiments
in \cite{suret2023soliton,fache2023interaction}.
See also \cite{el2021soliton,suret_soliton_2024}
for further theoretical and experimental developments.

The construction of SG kinetic equation in \cite{el2005kinetic} is also consistent with  the \textit{generalised hydrodynamics} (GHD) approach \cite{castro2016emergent, bertini2016transport, doyon2020lecture, spohn_hydrodynamic_2023}, 
which has proven to be a powerful theoretical framework 
for the understanding of large-scale, emergent hydrodynamic properties of integrable quantum and classical many-body systems. The relation between the KdV and Boussisnesq SGs, and their GHD description, was recently established in \cite{bonnemain2022generalized,bonnemain_soliton_2024}, 
where the formulation of their {\it thermodynamics} and generalized Gibbs ensembles via the thermodynamic Bethe ansatz was also worked out.

The collision rate ansatz and GHD approaches make it clear that the theory can be adapted to many soliton and interaction types \cite{doyon2017note}, including bidirectional dispersive wave systems exhibiting anisotropic soliton collisions, 
where phase shifts in overtaking and head-on interactions are different, e.g., see \cite{zhang_bidirectional_2003}. In this case one obtains a coupled system of two kinetic equations for different types of solitons \cite{congy2021soliton}. This has important implications for the behaviour of the physical wave field in a SG. Various physical systems exhibit anisotropic soliton interactions \cite{kaup_higher-order_1975, lee_resonant_2007}  whose generic properties are acutely captured by the classical integrable two-way Boussinesq equation  \cite{boussinesq1872theorie}. 
  
Virtually all works on SGs so far were limited to one spatial dimension, with the recent exception of \cite{leduque2024space}, where an experimental observation of 2D shallow water SG was reported.
In this paper we provide a first step towards the analytical description of the emergent hydrodynamics of a two-dimensional SG, by describing the {\it spatial statistics of a time-stationary but inhomogeneous gas}. 
We do so in the framework of the Kadomtsev-Petviashvili (KP)  equation \cite{kadomtsev1970stability},  the generalization of the KdV equation to weakly two-dimensional nonlinear dispersive waves \cite{ablowitz1981solitons} with applications to a broad variety of physical settings: from shallow water waves and plasma physics \cite{kodama_solitons_2013, Ruderman_2020} to ferromagnetism and Bose-Einstein condensates \cite{turitsyn_stability_1985, PhysRevA.67.023604, kamchatnov_stabilization_2008, hoefer_dark_2012}.   The KP equation arises in two versions, called KPI and KPII,
both integrable and possessing line soliton solutions \cite{biondini2003family,biondini2006soliton,biondini2007line,kodama2004young,kodama2017kp}; we concentrate on KPII, whose solitons are stable and which has broader physical relevance.

Our results are based on the observation that the Galilean-boosted KPII equation admits a time-stationary reduction that coincides with the integrable ``good'' Boussinesq equation \cite{mckean1981boussinesq} in which the Boussisneq time $y$ is one coordinate of the KP spatial $xy$-plane.
Trajectories of Boussinesq solitons correspond to KPII line solitons.
Hence a $(1+1)$D SG for the Boussinesq equation comprises a stationary (or propagating with constant velocity) $(2+0)$D line SG of the KPII equation: the space-time statistics of the Boussinesq SG gives the spatial statistics of the stationary KP line SG.
We leverage the recently developed kinetic theory of bidirectional anisotropic SGs \cite{congy2021soliton} and the GHD theory for the Boussinesq equation \cite{bonnemain_soliton_2024}, and infer some important implications for the respective 2D stationary gas of line KPII solitons.  Specifically, we introduce and study analytically the phenomena of the line soliton refraction by a 2D SG and of the spatial interference of two SGs.
We  verify the kinetic theory predictions numerically  by constructing the corresponding  KP and Boussinesq SGs via $N$-soliton solutions \cite{freeman1983soliton,biondini2006soliton,hirota2004direct,kodama2017kp} 
with large $N$  and appropriately chosen random distributions for the soliton parameters. Finally, we compute the correlations for stationary two-component KP SGs.

\medskip
\textbf{KP and Boussinesq equations, line soliton solutions and their interactions.}
The Galilean-boosted, dimensionless KPII equation reads 
\[
(u_t - u_x + 6 u u_x + u_{xxx})_x + u_{yy} = 0\,.
\label{e:KP}
\]
Time-independent solutions of the above equation satisfy the so-called ``good''  Boussinesq equation
\[
u_{yy} - u_{xx} + 3(u^2)_{xx} + u_{xxxx} = 0\,.
\label{e:Boussinesq}
\]
Eq.~\eqref{e:KP} admits a two-parameter family of line-soliton solutions
\[
u(x,y,t) = 2a^2\sech^2[a(x - c y - v t) + \theta_o]\,,
\label{e:KPsoliton}
\]
where $\theta_o$ is an arbitrary constant, and the soliton speed is completely determined in terms of the amplitude parameter~$a$ and the slope parameter $c$ by the soliton dispersion relation: $v= 4a^2 + c^2 - 1$. If $c = \pm \sqrt{1 - 4a^2}$, 
the above solution is stationary and is also a solution of 
the Boussinesq Eq.~\eqref{e:Boussinesq}.
This implies that:
(i) solitons of the Boussinesq equation cannot exceed the ``speed of sound'' $c=\pm1$ in either direction of propagation, and (ii) their amplitude must be smaller than $a = 1/2$, which is only reached when $c=0$. Note that the above relation determines the slope $c$ up to a sign, Eq.~\eqref{e:Boussinesq} is a bidirectional model.
In addition, both the KPII and Boussinesq equations admit a large family of (line-)soliton solutions,
describing elastic interactions among individual solitons.
These solutions, that we omit for brevity , are conveniently expressed via the Wronskian formalism
\cite{freeman1983soliton,biondini2006soliton,hirota2004direct,kodama2017kp,bogdanov2002boussinesq}.

Interaction between solitons of the KPII equation leads to a shift in the position of each soliton as $y\to\infty$ compared to its position as $y\to-\infty$,
and the same is true for the Boussinesq equation.
Let $\varphi_{n,m}$ be the phase shift of the $n$-th soliton 
as a result of the interaction with the $m$-th soliton.
Two kinds of soliton interactions can occur depending on the relative sign of
the slope of the individual solitons:
(i) co-directional (co),
when $c_n$ and~$c_{m}$ have the same sign,
and (ii) counter-directional (ct),
when $c_n$ and~$c_{m}$ have opposite sign. Though in both cases
\[
\label{e:phaseshifts}
\varphi_{n,m} = 
\frac{\sgn(a_{m}-a_{n})}{2}
    \ln \left[\frac{(c_n-c_{m})^2 - 12(a_n-a_{m})^2}
        {(c_n-c_{m})^2 - 12(a_n+a_{m})^2}\right],
\]
co- and counter-directional interactions are phenomenologically different. In particular, since soliton amplitudes are positive and bounded by $1/2$, co-directional shifts are always negative, $\varphi^{{\rm co}}_{n,m}<0$, implying solitons effectively repel each other. 
Conversely, counter-directional shifts $\varphi^{{\rm ct}}_{n,m}$ can either be positive or negative depending on the relative amplitudes of the two solitons.
This difference between the co- and counter-directional phase shifts is naturally interpreted as anisotropy of soliton interactions~\cite{congy2021soliton}. 
Note that the logarithms in Eq.~\eqref{e:phaseshifts} implies that $N$-soliton solutions may develop resonances at finite values of $y$ for certain sets of amplitudes. We restrict ourselves to regular solutions; conditions ensuring regularity at all time were first introduced in \cite{lambert1987soliton} and again discussed in \cite{bonnemain_soliton_2024}. Those conditions imply that no two solitons can have the same amplitude and that $\varphi_{n,m}\neq0$ for admissible non-zero amplitudes.

\begin{figure*}[t!]
 \unitlength=1cm
 \centerline{\includegraphics[width=0.32\textwidth]{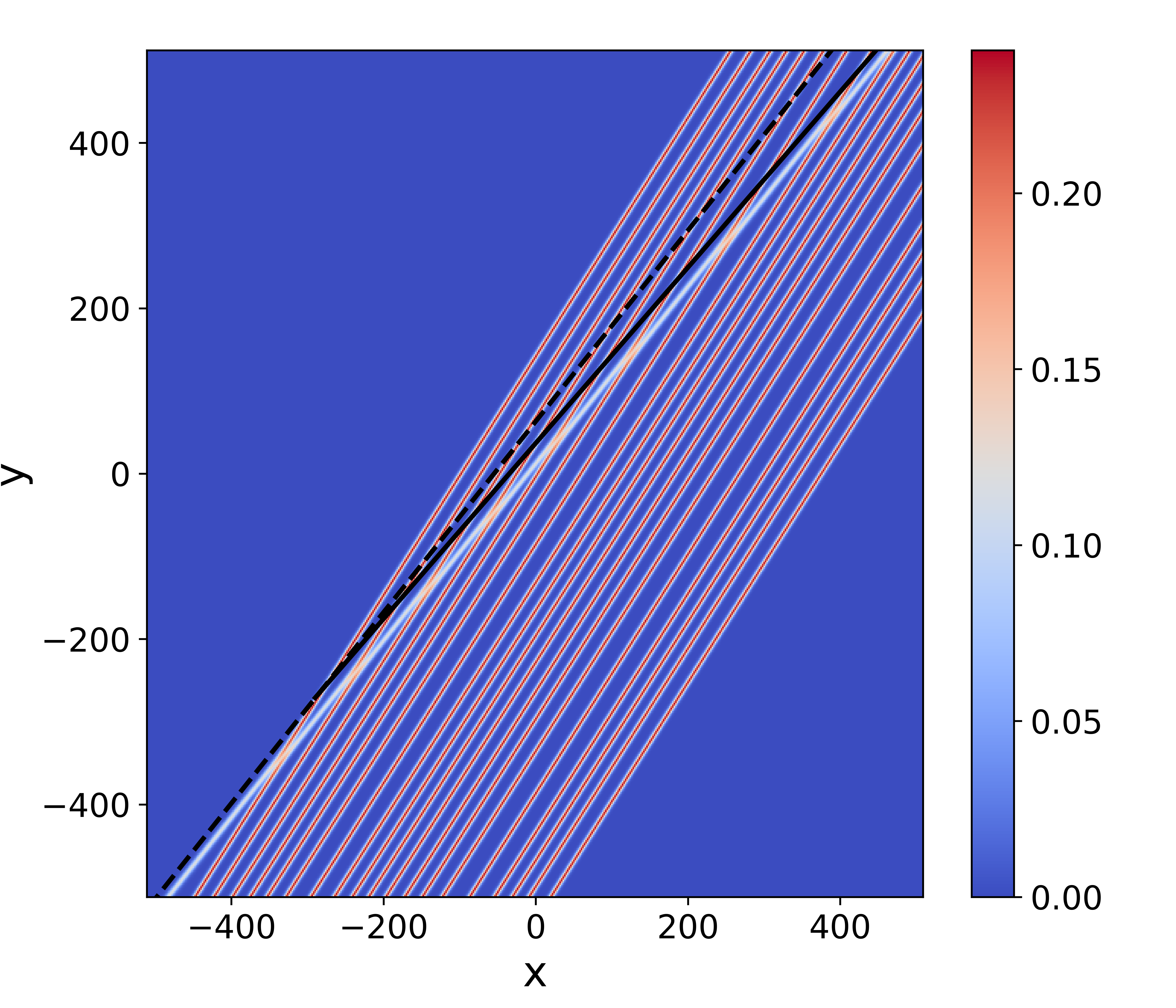}
 \includegraphics[width=0.32\textwidth]{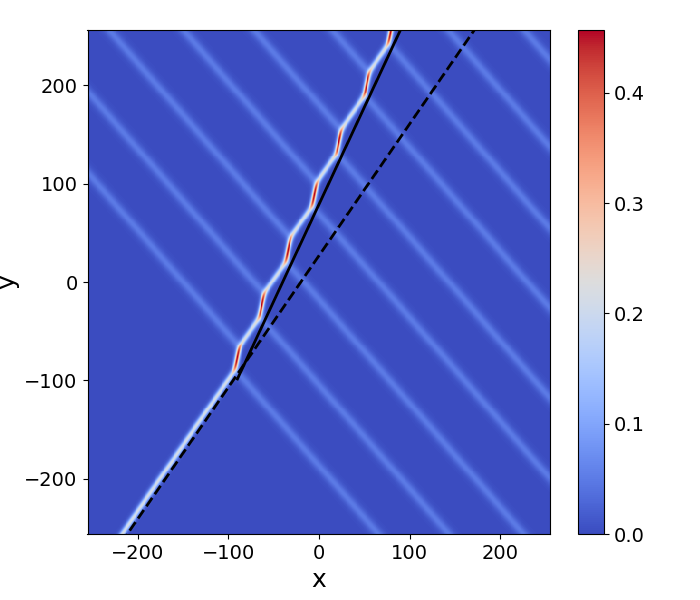}
\includegraphics[width=0.32\textwidth]{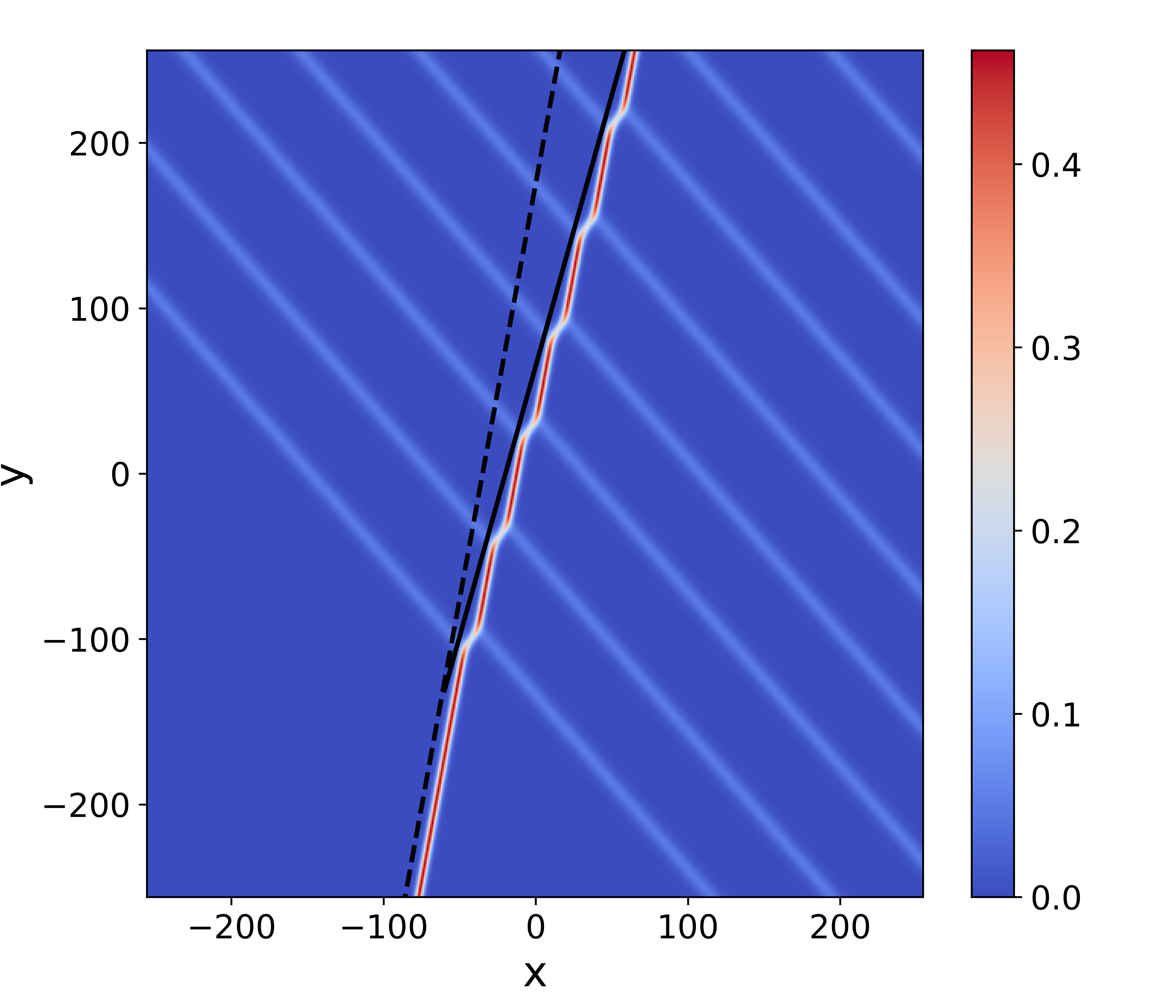}}
 \caption{Refraction of a trial line soliton by a monochromatic 2D stationary KPII SG composed of $N_g=19$ solitons. 
 Left: co-directional, r-r refraction of a trial soliton with spectral parameter $\ar_1=0.25$ by a monochromatic 2D stationary SG with spectral parameter $\ar_2=0.366$, density $\Fr_2=0.04$ and $\delta_g=1.\times10^{-3}$. 
 Centre and right: Counter-directional, r-l refraction of a trial line soliton by a monochromatic 2D stationary SG with spectral parameter $\al_2=0.158$, density $\Fl_2=0.013$ and $\delta_g=1.\times10^{-6}$. 
 The spectral parameters of the trial soliton are $\ar_1=0.49$ (middle panel, negative refraction) and  $\ar_1=0.3318$  (right panel, positive refraction). The black dashed lines represent the slope of a non-interacting line soliton, and the black solid lines represent the analytically predicted effective slope of the trial r-soliton resulting from the cumulative phase shift in the interaction with the SG.}
 \label{fig:OT_HO}
 \end{figure*}

\medskip
\textbf{Kinetic equation for a 2D stationary SG.}
Following the general theory of bidirectional SGs \cite{congy2021soliton} we introduce the separate DOS's $\fl(a;x,y)$ and $\fr(a;x,y)$  for the negatively (left)- and positively (right)-oriented line solitons (referred in what follows as r-solitons and l-solitons). At ``time''  $y$, $f^{\alpha}(a;x,y) \dd a  \dd x$ represents the  number of $\alpha$-solitons ($\alpha$ being either l or r) in the element $[a,a+ \dd a]\times[x,x+ \dd x]$ of the phase space $A^{\alpha} \times \mathbb R$, $A^\alpha$ being the ``spectral support'' of the respective DOS under conventional SG terminology \cite{el2021soliton}. We should stress that the $(x,y)$-variations of the total DOS, $f(a;x,y)=\fl(a;x,y) +\fr(a;x,y)$, in a non-homogeneous SG, occur on a hydrodynamic scale much larger than the ``mesoscopic'' scale where the gas is locally homogeneous, itself much larger than the microscopic scale of variations of the wave field (the coherence length), see \cite{suret_soliton_2024}.

The collision rate ansatz  of \cite{congy2021soliton}  applied to the Boussinesq equation yields the kinetic equation for the steady-state weakly non-uniform KPII SG in the form of two continuity equations  describing the DOS's variations in $(x,y)$-plane: 
\vspace*{-0.5ex}
\begin{equation}
\label{e:RarContEq}
\fl_y + (\ceffl \fl)_x =0, \quad
\fr_y + (\ceffr\fr)_x = 0,
\end{equation}
where $s^{\rm l,r}(a;x,y)$ are the effective  slope parameters 
satisfying the coupled integral equations of state, dropping the $(x,y)$-dependence for brevity
\begin{multline}
\label{e:Ceff}
    s^\alpha(a) = c^\alpha(a) + 
    \\
    \sum_{\gamma = \alpha, \beta}\int_{A^\gamma}\Delta(a,a')f^\alpha(a')\left[s^\alpha(a)-s^\gamma(a')\right]\dd a' ,
\end{multline}
where $\Delta(a,a') = \varphi(a,a')/a$ is the position shift, $\alpha$ is either ${\rm l}$ ($c^l(a)=-\sqrt{1-4a^2}$) or r ($c^{\rm r}(a)=\sqrt{1-4a^2}$), $\beta$ either r or l respectively. 
The equations of state \eqref{e:Ceff} are consistent with the Euler GHD of the Boussinesq SG constructed in \cite{bonnemain_soliton_2024}. 

Following \cite{el2005kinetic, el2011kinetic, congy2021soliton, congy2024riemann} we introduce the ``polychromatic'' delta-functional ansatz for both species' DOS:
\vspace*{-1ex}
\begin{equation}\label{polychrom}
\begin{aligned}
&\fr(a;x,y) = \sum_{i=1}^{M^{\rm r}} \Fr_i(x,y)\delta(a-\ar_i) \;,
\\
&\fl(a;x,y) = \sum_{i=M^{\rm r}+1}^{M^{\rm r}+M^{\rm l}} \Fl_i(x,y)\delta(a-\al_i) \;,
\end{aligned}
\end{equation}
where $M^{\rm l,r}$ are the numbers of components of the respective soliton species. As a matter of fact, this polychromatic reduction is an idealisation, and in practice, amplitudes are considered to be uniformly distributed around $a_i^\alpha$ in small intervals of size $\epsilon_g$. Substituting~\eqref{polychrom} in the kinetic equations \eqref{e:RarContEq} and~\eqref{e:Ceff} yields a system of hydrodynamic conservation laws for $\Fl_i(x,y)$ and $\Fr_i(x,y)$:
\begin{equation}\label{e:RarContEq_Ht}
   (\Fr_i)_y +  (\ceffr_i \Fr_i )_x =0, \quad
    (\Fl_j)_y +  (\ceffl_j\Fl_j)_x = 0,
\end{equation}
$i=1, \dots M^{\rm r}$, $j=M^{\rm r}+1, \dots, M^{\rm r}+M^{\rm l}$,
complemented by a system of algebraic closure equations connecting the components' slope parameters $\ceffl_i\equiv \ceffl(\al_i), \ \ceffr_i\equiv \ceffr(\ar_i)$ with the densities $\Fl_i, \ \Fr_i$, which are omitted here for brevity. 
The hydrodynamic reductions of the kinetic equation were shown in \cite{el2011kinetic} to be linearly degenerate integrable systems of hydrodynamic type, which implies the absence of wave breaking and the admissibility of contact discontinuities in solutions of canonical Riemann problems \cite{el2005kinetic, congy2021soliton, congy2024riemann}.

Here we consider two-component, ``bi-chromatic'', SGs with $M^{\rm l} + M^{\rm r}=2$ to describe the following prototypical interaction problems:
(i) 2D  refraction of a line soliton by a 1-component (monochromatic) SG; 
(ii) oblique interference of two monochromatic SGs. 
Each of these problems includes two qualitatively different configurations: a) co-directional refraction/interference when both the trial soliton and the SG in Problem (i) or two monochromatic SGs in Problem (ii) have the same orientation; and b) counter-directional refraction/interference when the trial soliton and the SG or two interacting SGs have opposite orientations. We shall use the shorthand notation ``r-r''  for the two-component interaction with both gas components having right orientation and ``r-l'' if the 1st component has right orientation and the 2nd--left orientation etc. Moreover, in the context of the polychromatic reduction we shall use the notation $c^{\alpha}_i\equiv c^{\alpha}(a^{\alpha}_i)$.

\begin{figure}[t!]
 \centerline{\includegraphics[width=0.245\textwidth]{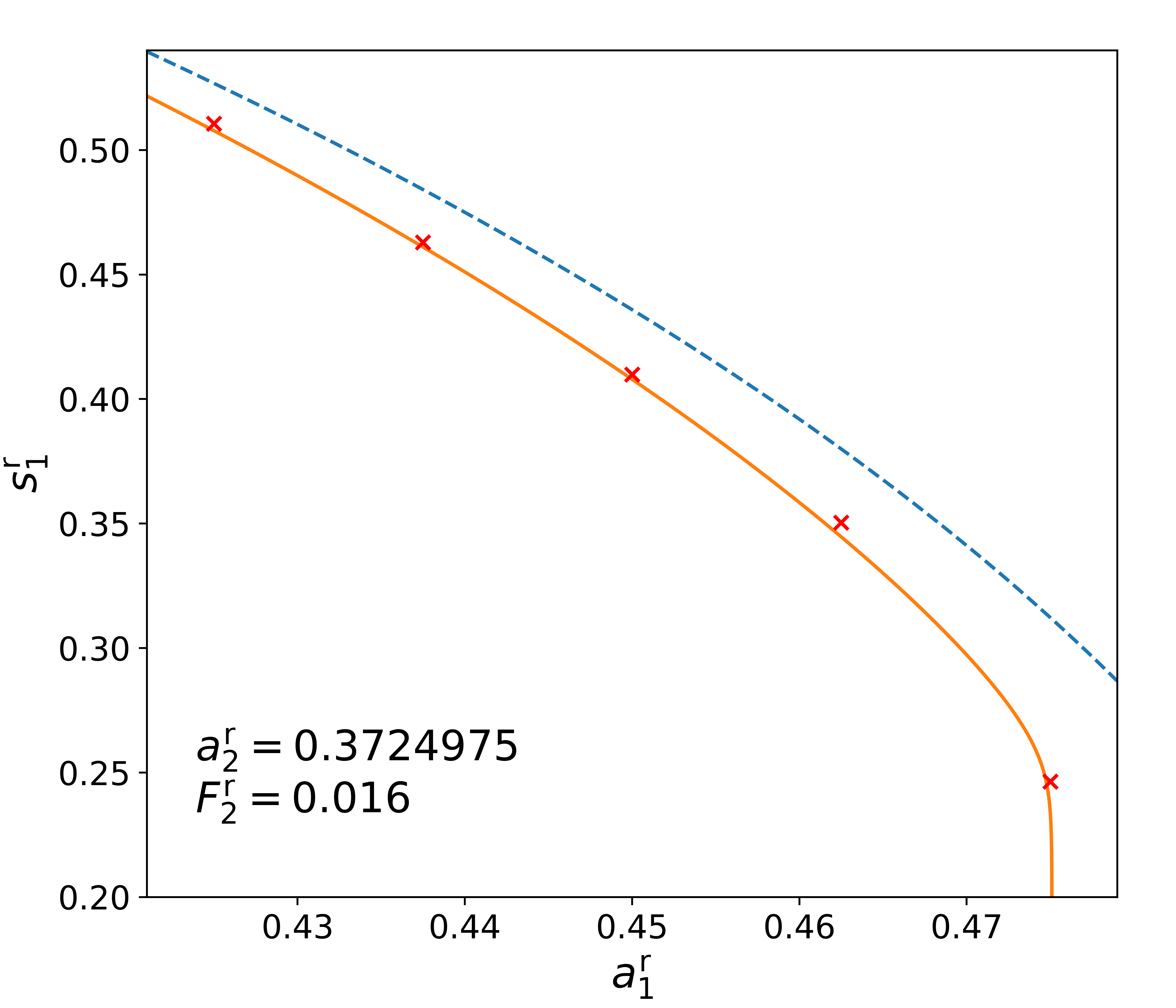}
 \includegraphics[width=0.245\textwidth]{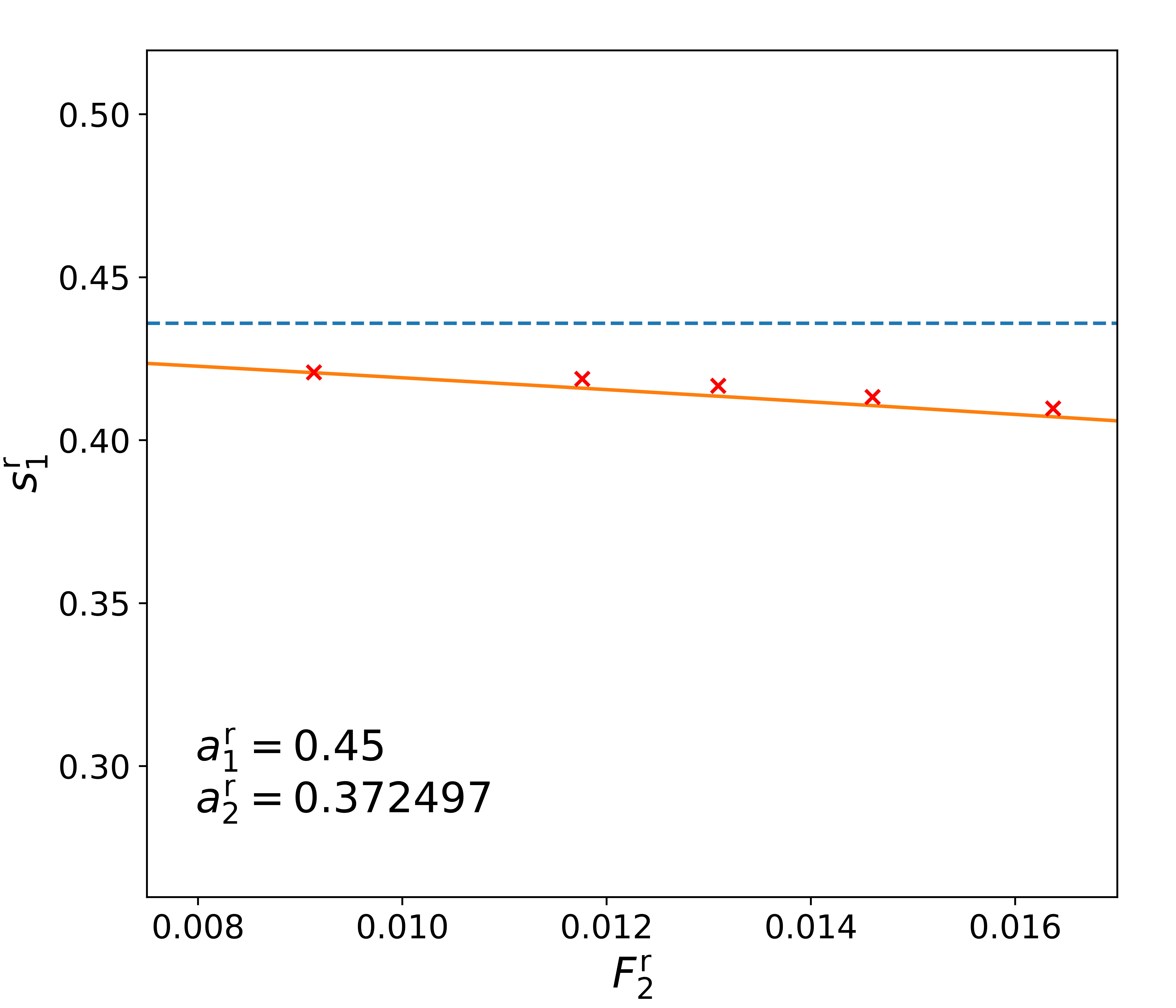}}
 \centerline{\includegraphics[width=0.245\textwidth]{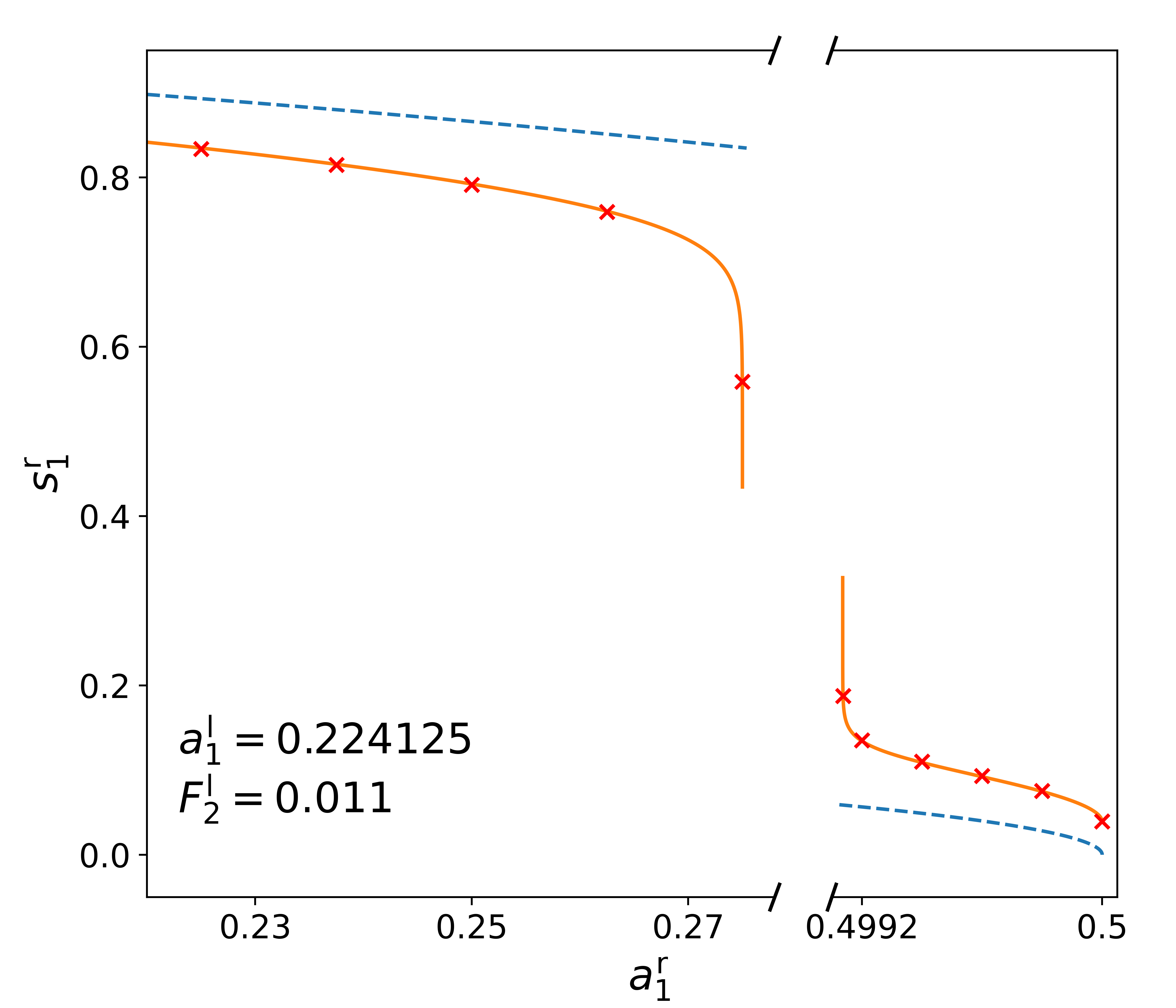}
 \includegraphics[width=0.245\textwidth]{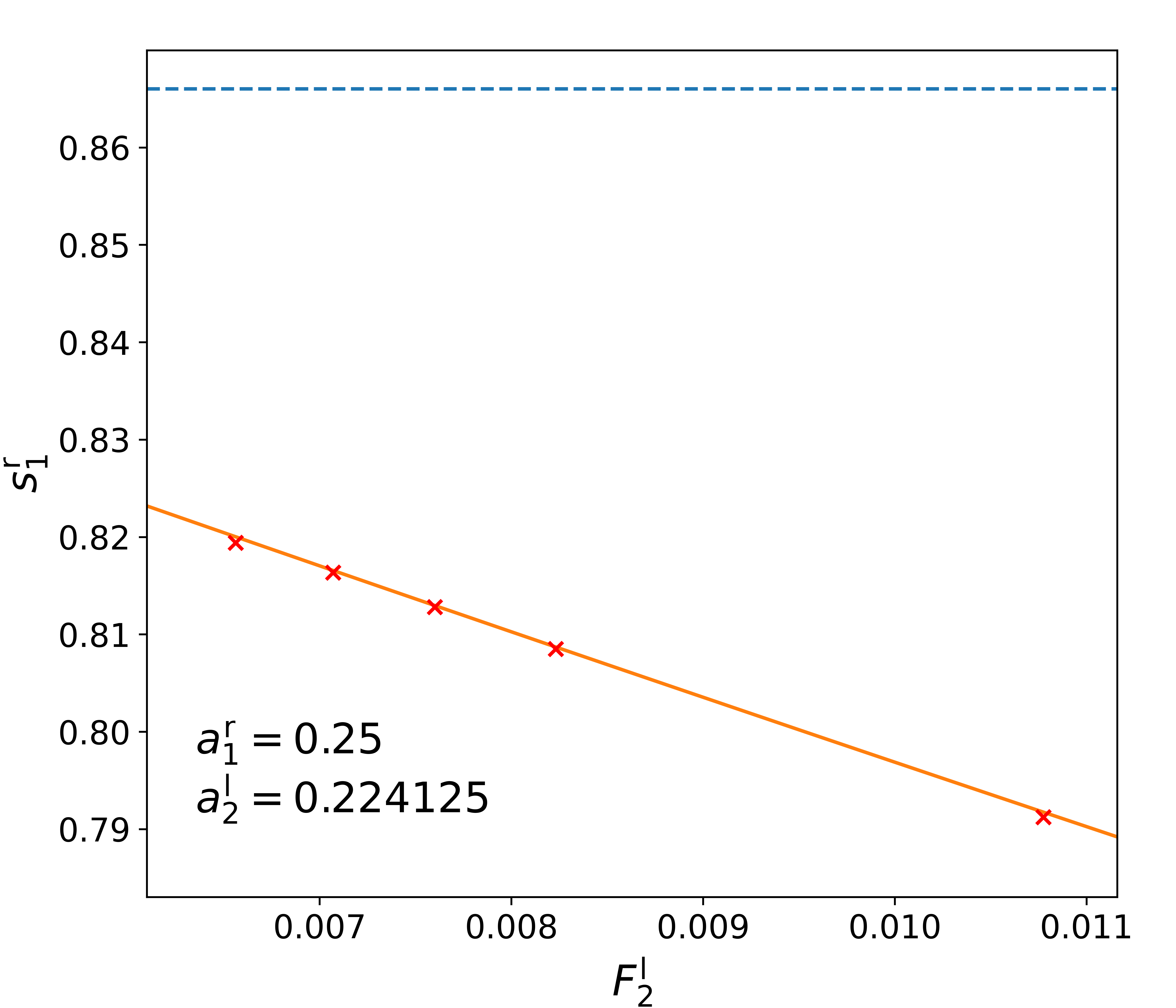}}
 \caption{Effective slope $s^{\rm r}_1$ for the co-directional (top) and counter-directional (bottom) refraction of a soliton of parameter $\ar_1$ by a  monochromatic SG of parameters $a^{\alpha}_2$ and $F^{\alpha}_2$. The SG is composed of $N=100$ solitons with $\delta_g=2.5\times10^{-9}$, see Appendix for details. 
 Comparison between the theoretical prediction (solid line) and numerics (red cross). The dashed curves show the slope of a non-interacting trial line soliton.}
 \label{fig:eff_vel_OT}
 \end{figure}

(i.a) {\it Co-directional r-r refraction}. (The l-l case is analogous). In this case $M^{\rm r}=2$, $M^{\rm l}=0$. To describe the refraction of a single r-soliton of amplitude $\ar_1$  by the right-oriented monochromatic gas  with the amplitude $\ar_2>\ar_1$ ($\crr_2<\crr_1$) and  density $\Fr_2$, we set $\Fr_1 =0$ in the two-component r-r reduction of the equations of state \eqref{e:Ceff} so that the effective slope parameter of the refracted soliton is given by (cf. \cite{congy2021soliton} and Appendix)
\bse
\label{refractionslope}
\begin{equation}
\label{codir_ref}
\ceffr_1 =\crr_{1}+\frac{\big[\crr_{1}- \crr_{2}\big]\Delta^{\rm co}(\ar_1,\ar_2)\Fr_{2}}{1-\Delta^{\rm co}(\ar_1,\ar_2)\Fr_{2}}
\end{equation}
Since $\Delta^{\rm co} <0$, it follows from \eqref{codir_ref} that 
$\sgn [\ceffr_1 - \crr_{1}] =-1$, i.e. the co-directional refraction is always positive, according to the conventional ray optics terminology. 
See Fig.~\ref{fig:OT_HO}(left) for a typical co-directional refraction configuration.

(i.b) {\it Counter-directional r-l refraction}.  We set $M^{\rm r}=M^{\rm l}=1$ so that the equations of state \eqref{e:Ceff} yield, upon further setting  
$\Fr_1 = 0$, the effective slope parameter for the refracted r-soliton:
\begin{equation}\label{ctdir_ref}
\ceffr_1 =\crr_{1}+ \frac{\big[\crr_{1}- \cl_{2}\big]\Delta^{\rm ct}(\ar_1,\al_2)\Fl_{2}}{1-\Delta^{\rm ct}(\ar_1,\al_2)\Fl_{2}}  .  
\end{equation}
\ese
Since $(1-\Delta^{\rm ct}(\ar_1,\al_2)\Fl_{2})>0$, $\sgn [\ceffr_1 - \crr_{1}] =  \sgn\big(\Delta^{\rm ct}(\ar_1,\al_2)[\crr_{1}-\cl_{2}]\big)$ can be either positive or negative. See Fig.~\ref{fig:OT_HO}  for the numerical demonstration of the negative (middle panel) and positive (right panel) counter-directional 2D soliton refraction.

Comparison of the analytically determined effective refraction slope parameters \eqref{refractionslope} with the results of direct numerical implementation of the 2D line soliton refraction problems for KPII, in Fig.~\ref{fig:eff_vel_OT}, demonstrates excellent agreement.
Note that ``temporal'' refraction of an optical soliton interacting with a 1D NLS equation SG has been observed in fiber optics experiment \cite{suret2023soliton}.

\begin{figure}[t!]
 \includegraphics[width=0.48\textwidth]{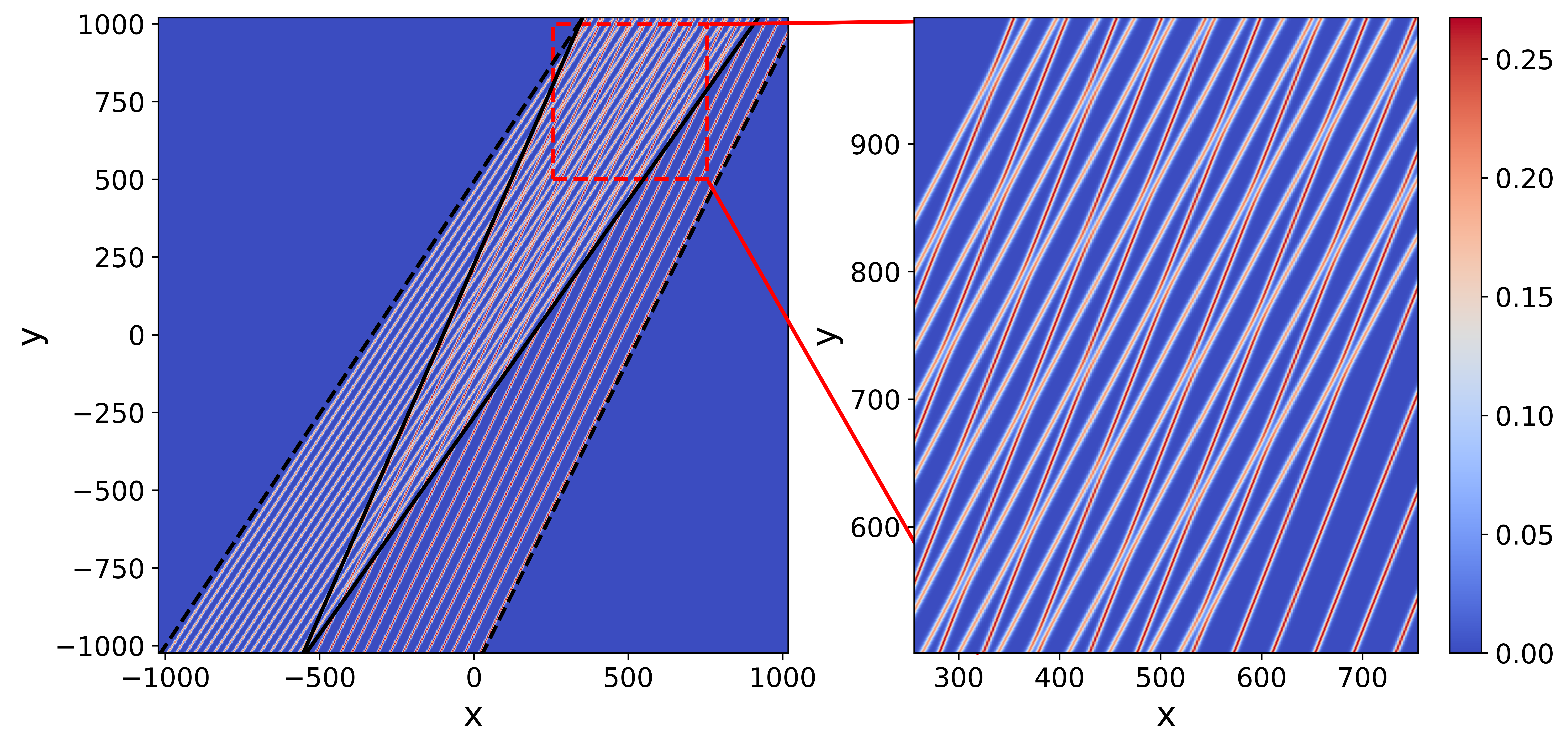}
 \includegraphics[width=0.48\textwidth]{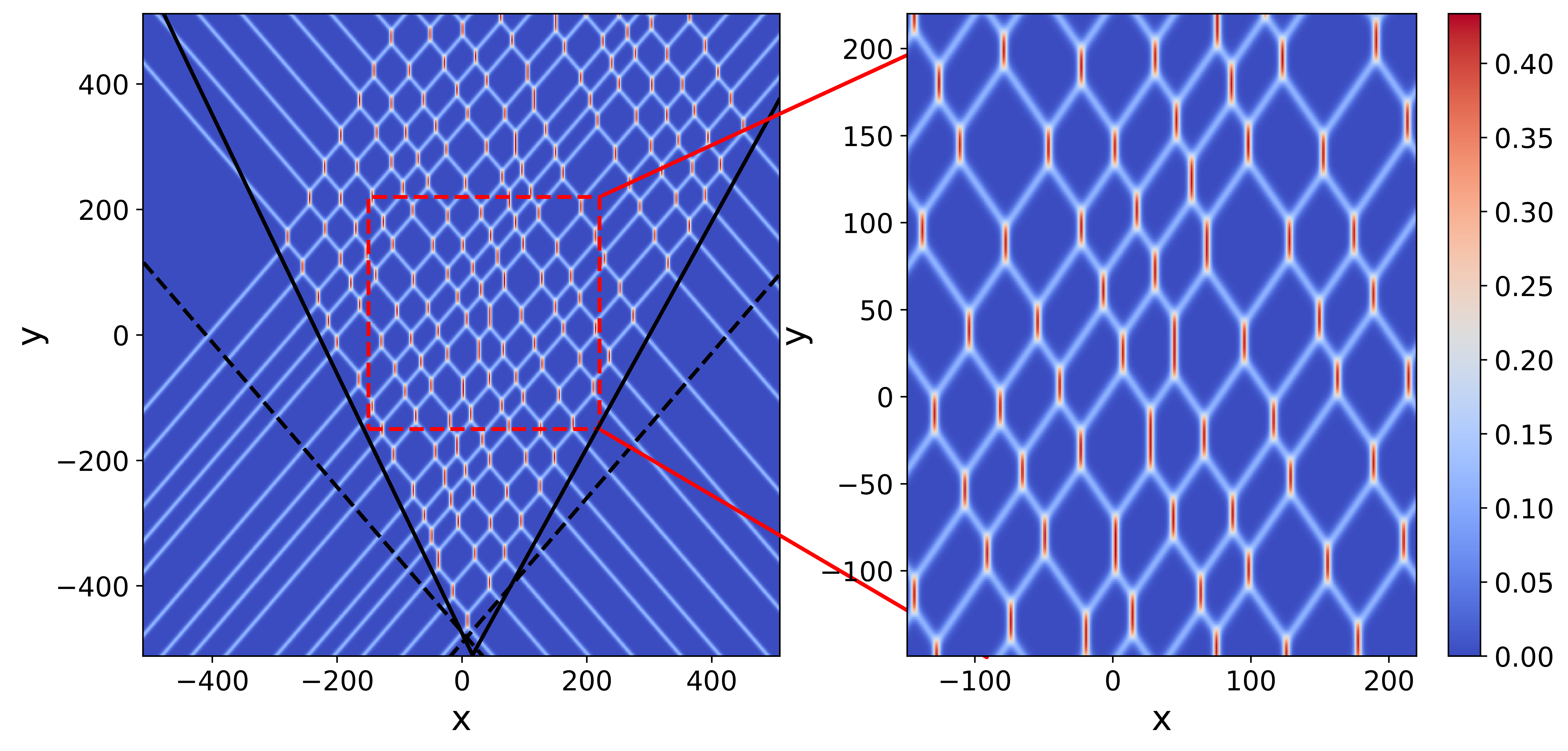}
\caption{Oblique nonlinear interference of two stationary monochromatic SGs for the KPII equation. (top) co-directional interference. Parameters: $\ar_1=0.3715$, $\Fr_1=0.036$, $\ar_2=0.433012$, $\Fr_2=0.028$, $N^{\rm r}_1=N^{\rm r}_2=15$ and $\delta_g = 1.0\times 10^{-4}$. (bottom) counter-directional interference. Parameters: $\ar_1=0.25$, $\Fr_1=0.017$, $\al_2=0.25$, $\Fl_2=0.014$, $N^{\rm r}_1=N^{\rm l}_1=15$ and $\delta_g = 1.0\times 10^{-4}$. See Appendix for details. 
The dashed lines show the slope parameters of the monochromatic SGs outside the interaction region; the black solid lines delimit the boundaries of the nonlinear interference region defined by the slopes of hydrodynamic contact discontinuities.} 
\label{fig:HC_rg}
\end{figure}

\medskip
The {\it oblique interference} of two monochromatic gases is described by the solution of the  Riemann problem for the hydrodynamic system \eqref{e:RarContEq_Ht} with $M^l + M^r =2$, specifying a one-component density at each side of $x=0$ at $y=0$. Due to linear degeneracy of polychromatic reductions of the kinetic equation \cite{el2011kinetic} the weak solution consists of three constant states for each of the components, separated by two oblique contact discontinuities whose slope parameters $c^+>c^-$ are found from the Rankine-Hugoniot conditions (cf.~\cite{el2005kinetic, congy2021soliton} and Appendix):

\bse 
\label{contact}
\begin{align}
c^+= c^\alpha_{1} +\frac{\big[c^\alpha_{1}-c^\beta_2\big]\Delta_{1,2}F^\beta_{2}}{1-\Delta_{2,1}F^\beta_{2}}, \label{contact_a} 
\\
c^-=c^\beta_2-\frac{\big[c^\alpha_{1}-c^\beta_2\big]\Delta_{2,1}F_{1}^\alpha}{1-\Delta_{1,2}F^\alpha_1},\label{contact_b}
\end{align}
\ese
where $\Delta_{i,j}=\Delta\left(a^{\alpha}_i,a^{\beta}_j\right)$.  For co-directional r-r oblique interference, $\alpha=\beta=r$; 
for r-l interference, $\alpha=r$, $\beta=l$. As one can see, the leading contact discontinuity slope \eqref{contact_a} coincides with the refraction slope of a single line soliton for co-directional \eqref{codir_ref} and for counter-directional \eqref{ctdir_ref} interference. 

Two prototypical oblique nonlinear interference configurations are illustrated in Fig.~\ref{fig:HC_rg}; the positions of the contact discontinuities, \eqref{contact_a} and \eqref{contact_b}, are shown in black solid lines and display excellent agreement with numerical results. 
Nonlinear soliton interaction widens the interference region, while in the counter-directional case the effect is opposite, consistent with the signs of the respective two-soliton phase shifts \eqref{e:phaseshifts}. One can also observe the qualitative difference in the geometry of the interference patterns with the characteristic ``honeycomb SG'' signature in the counter-directional case.

\textbf{Correlations.}
We now consider the spatial correlations for the 2D bi-chromatic SGs. 
These can be evaluated by mapping the GHD results for Boussinesq SGs \cite{doyon2018exact,bonnemain_soliton_2024}, to the (2+0)D KPII setup. Within any region of $(x,y)$ where the SG is  statistically homogeneous (e.g in the honeycomb SG region in the bottom-left panel of Fig.~\ref{fig:HC_rg}), we have~
\begin{equation}
\label{bicov}
    \int_{\mathbb R} \dd x \,\langle u(x,y)u(0,0) \rangle^c = 16 \big[(a_2^\beta)^2F_1^\alpha  + (a_1^\alpha)^2F_2^\beta\big] \; ,
\end{equation}
$(\alpha,\beta)$ being defined as in Eqs.~\eqref{contact}, $\langle u(x,y)u(0,0) \rangle^c = \langle u(x,y)u(0,0) \rangle - \langle u(x,y)\rangle\langle u(0,0) \rangle$ being the connected correlation. Note that the right-hand side does not depend on $y$, because the field $u$ is a conserved density of the Boussinesq equation. 
Nontrivial two-point correlations arise due to the inevitably present physical spectral broadening $\sim \epsilon_g$ around the points $a_i^\alpha$ in a polychromatic gas with the DOS \eqref{polychrom} .
Then  along a given effective slope direction (e.g. $s_1^\alpha$) we obtain 
\vspace*{-1ex}
\begin{multline}
    \frac{1}{2L}\int_{-L}^L \dd x \langle u(s_1^\alpha Y +x,Y)u(0,0) \rangle^c 
    \sim \frac{16 (a^\beta_2)^2F_1^\alpha}{Y\epsilon_g |\partial_{a_1^\alpha}{s_1^\alpha}|} 
\label{bitwopoint}
\end{multline} 
for $L\gg \max\left[\delta x(Y,\epsilon_g), \ell\right]$, and $Y\gg L$. Here $\ell = \left[F_1^\alpha +F_2^\beta \right]^{-1}$ quantifies the average inter-soliton spacing along the $x$~direction, and $\delta x(Y,\epsilon_g)$ is the size of the interval around $x=s_1^\alpha Y$ at which the soliton would intersect with the horizontal, given $\epsilon_g$, at $y=Y$  (the result should be independent of the precise choice of $L$). Note that for a fixed  $\epsilon_g>0$ these correlations exhibit the $1/y$ ballistic scaling typical of many-body integrable systems \cite{doyon2020lecture}.

\textbf{Numerical implementation.}
The numerical implementation of 2D  SGs studied in this paper is achieved via exact multi-soliton solutions of the KPII equation configured accordingly to the designed DOS. The numerical algorithm relies on the Wronskian formulation of the $N$-soliton solution, and high-precision arithmetic routines are included to overcome the numerical accuracy problems arising for $N\gg 1$ .

\medskip
\textbf{Discussion.}
Summarising,  we have constructed kinetic theory of 2D stationary SGs for the KPII equation. This has been done using the steady-state reduction of the  KPII equation to the (1+1) integrable two-way Boussinesq equation and applying the recently developed kinetic theory of bidirectional SGs \cite{congy2021soliton} and the GHD of the Boussinesq equation \cite{bonnemain_soliton_2024}. Using the polychromatic, delta-function reductions of the spectral kinetic equation for the Boussinesq equation we have considered two basic 2D SG interaction problems: (i) refraction of a line soliton by a spectrally ``monochromatic'' SG; and (ii) oblique interference of two SGs described by weak solution of the stationary Riemann problem.  The phenomena of 2D soliton refraction and SG interference can be potentially realised in a 2D shallow water tank, e.g. using the setting similar to \cite{leduque2024space}. The oblique SG interference can also be used for the modelling of the crossing sea type phenomena in the ocean \cite{pelinovsky_non-gaussian_2016}.  

Analytical predictions regarding the 2D SG interactions are shown to be in excellent agreement with the direct numerical implementation of  2D stationary SGs using appropriately configured exact $N$-soliton solutions of the KPII equation.   These results represent the first step in the development of the theory of 2D SGs. 
A natural next step will be the formulation of an analytical description of non-stationary SGs of the full KPII equation which will include a variety of physical phenomena such as soliton resonance and web structure \cite{biondini2003family,biondini2006soliton,biondini2007line,kodama2004young,kodama2017kp} not accounted for in the stationary reduction.

\medskip
\textbf{Acknowledgements.}
The authors thank the Isaac Newton Institute for Mathematical Sciences, Cambridge, for support and hospitality during the programme ``Emergent phenomena in nonlinear dispersive waves'', where work on this paper was undertaken. This work was supported by Engineering and Physical Sciences Research Council (EPSRC) grant EP/R014604/1. TB and BD were further supported by the EPSRC under grant EP/W010194/1. The authors thank R. Dubertrand for sharing C. Schmit's algorithm for efficiently computing matrix determinants.

%%%%%%%%%%%%%%%%%%%%%%%%%%%%%%%%%%%%%%%%%%%%%%%%%%%%%%%%%%%%%%%%%%%%%%%%%%%%%%%%%%%%%%%%%%%%%%%%%%%
%%%%%%%%%%%%%%%%%%%%%%%%%%%%%%%%%%%%%%%%%%%%%%%%%%%%%%%%%%%%%%%%%%%%%%%%%%%%%%%%%%%%%%%%%%%%%%%%%%%
%%%%%%%%%%%%%%%%%%%%%%%%%%%%%%%%%%%%%%%%%%%%%%%%%%%%%%%%%%%%%%%%%%%%%%%%%%%%%%%%%%%%%%%%%%%%%%%%%%%

\def\o#1{^{(#1)}}
\newcommand\partialderiv[3][]{\frac{\partial^{#1}#2}{\partial {#3}^{#1}}}
\let\be=\[
\let\ee=\]
\def\Wr{\mathop{\rm Wr}\nolimits}
\def\e{{\mathrm{e}}}
\def\half{{\textstyle\frac12}}
\def\diag{\mathop{\rm diag}\nolimits}
\def\txtfrac#1#2{{\textstyle\frac{#1}{#2}}}

%%%%%%%%%%%%%%%%%%%%%%%%%%%%%%%%%%%%%%%%%%%%%%%%%%%%%%%%%%%%%%%%%%%%%%%%%%%%%%%%%%%%%%%%%%%%%%%%%%%
%%%%%%%%%%%%%%%%%%%%%%%%%%%%%%%%%%%%%%%%%%%%%%%%%%%%%%%%%%%%%%%%%%%%%%%%%%%%%%%%%%%%%%%%%%%%%%%%%%%
%%%%%%%%%%%%%%%%%%%%%%%%%%%%%%%%%%%%%%%%%%%%%%%%%%%%%%%%%%%%%%%%%%%%%%%%%%%%%%%%%%%%%%%%%%%%%%%%%%%

\appendix

\section*{Appendix}

%\bigskip
\subsection{Multi-soliton solutions}
It is easy to see that the Galilean-boosted KPII equation~\eqref{e:KP}, i.e., the PDE
\vspace*{-1ex}
\[
(u_t - u_x + 6 u u_x + u_{xxx})_x + u_{yy} = 0\,,
\label{e:KP2}
\]
can be obtained from the standard KPII equation via the change of 
independent variables $x' = x - t$,
$y'=y$ and $t'=t$.
Accordingly, Eq.~\eqref{e:KP2}
admits a large class of solutions expressed via the Wronskian formalism
\cite{biondini2006soliton,freeman1983soliton,hirota2004direct,kodama2017kp}:
\vspace*{-1ex}
\[
u(x,y,t) = 2\partialderiv[2]{}x\log\tau(x,y,t)\,,
\label{e:utau}
\]
where the ``tau'' function $\tau(x,y,t)$ satisfies Hirota's bilinear equation \cite{hirota2004direct}.
It was then shown in~\cite{freeman1983soliton} that solutions of Hirota's bilinear equation can be obtained in terms of Wronskians:
\vspace*{-1ex}
\begin{multline}
\tau(x,y,t) = \Wr(f_1,\dots,f_N)
\\
 = \det\begin{pmatrix} 
f_1 & f_2 & \cdots & f_N \\
f_1\o1 & f_2\o1 & \cdots & f_N\o1 \\
\vdots & \vdots & \ddots & \vdots \\
f_1\o{N-1} & f_2\o{N-1} & \cdots & f_N\o{N-1} 
\end{pmatrix},
\label{e:Wronskian}
\end{multline}
where $f_n\o{j} = \partial^j f/\partial x^j$, and 
$f_1,\dots,f_N$ satisfy the linear PDEs
\[
\partialderiv{f_n}y = \sqrt3\,\partialderiv[2]{f_n}x\,,\quad
\partialderiv{f_n}t = \partialderiv{f_n}x - 4\partialderiv[3]{f_n}x\,.
\label{e:fPDEs}
\]
Multi-soliton solutions of~\eqref{e:KP2} are obtained by choosing 
$f_1,\dots,f_N$ as linear combinations of exponentials: 
\vspace*{-1ex}
\be
f_n(x,y,t) = \sum_{m=1}^M g_{n,m} \e^{\theta_m}\,,
\label{e:fexponentials}
\ee
where, similarly to~\cite{biondini2006soliton}, the exponential ``phases'' $\theta_1,\dots,\theta_M$ are given by
\be
\theta_m = k_m x + \sqrt3 k_m^2 y + (k_m - 4k_m^3)t + \theta_{m,0}\,.
\ee
The above solution is uniquely determined by the ``phase parameters'' $k_1,\dots,k_M$ and the $N\times M$ coefficient matrix $G = (g_{n,m})$,
plus the ``translation constants'' $\theta_{1,0},\dots,\theta_{M,0}$.
Without loss of generality, one can take the phase parameters to be ordered so that $k_1<\cdots<k_M$.

The simplest nontrivial case is obtained when $N =1$ and $M=2$.  
In this case one recovers the soliton solution~\eqref{e:KPsoliton} of the KPII equation. It is convenient to label the two phase parameters as $k_-$ and $k_+$ in this case.
The corresponding amplitude and slope parameters $a$ and $c$ are then given by the map
\vspace*{-1ex}
\bse
\begin{equation}
a = \half\,(k_+ - k_-)\,,\qquad
c = - \sqrt3\,(k_+ + k_-)\,,
\label{e:directmap}
\end{equation}
which is inverted by
\vspace*{-1ex}
\begin{equation}
k_\pm = \pm a - c/(2\sqrt3)\,.
\label{e:inversemap}
\end{equation}
\ese
For the Boussinesq equation~\eqref{e:Boussinesq}, since $a$ is fixed by the dispersion relation 
\be
c = \pm \sqrt{1 - 4a^2}
\label{e:Bamplitude},
\ee
$k_\pm$ can instead be expressed in terms of the single parameter $c$ as
\be
k_\pm = \half\,\big(\pm \sqrt{1-c^2} - c/\sqrt3\big)\,.
\ee

For arbitrary values of $N$ and $M$, the Wronskian representation~\eqref{e:Wronskian} 
with $f_1,\dots,f_N$ as in~\eqref{e:fexponentials}
yields
\begin{multline}
\tau(x,y,t)=  \det(K\,\e^{\Theta}G^T)\\
 = \kern-1em\sum_{1\le m_1<m_2<\dots<m_N\le M} \kern-1em
  V_{m_1,\dots,m_N}\,
  G_{m_1,\dots,m_N}\,\e^{\theta_{m_1}+\cdots+\theta_{m_N}}\,,
\\
\label{e:tauNM}
\end{multline}
where $\Theta=\diag(\theta_1,\dots,\theta_M)$, 
the $N\times M$ matrix $K$ is given by $K=(k_m^{n-1})$ 
$V_{m_1,\dots,m_N}>0$ is a Van~der~Monde determinant,
and $G_{m_1,\dots,m_N}$ is the determinant of the $N\times N$ minor obtained from 
columns~$m_1,\dots,m_N$ of {$G$.
Note that the spatio-temporal dependence of the tau-function is
confined to the exponential phases.

For irreducible coefficient matrices,
the above representation produces a solution with exactly $N$ asymptotic line solitons as $y\to\infty$
and $M-N$ asymptotic line solitons as $y\to-\infty$
\cite{biondini2006soliton}.
The amplitude and slope of each asymptotic line soliton are completely determined by a pair of phase parameters among $k_1,\dots,k_M$, 
but the precise pair associated to each soliton depends on the choice of coefficient matrix~$G$.
Importantly, in order for solutions to be non-singular, all the $N\times N$ minors of the associated coefficient matrix $G$ are non-negative.

A Wronskian formulation for the multi-soliton solutions of the Boussinesq equation was given in~\cite{nimmo1983method}, 
but it can be immediately obtained as a reduction of the above formalism for the KPII equation,
except that one has additional restrictions on the choice of phase parameters,
as we discuss next.
Specifically, 
suppose that one wants to construct a multi-soliton solution consisting of two line solitons with 
amplitude and slope parameters $(a_1,c_1)$ and $(a_2,c_2)$. 
One obviously needs $N=2$ and $M=4$.
Denote the two sets of phase parameters, one set associated to each soliton, as $k_{1,\pm}$ and $k_{2,\pm}$, respectively.
The resulting two-soliton solution differs
depending on the relative ordering of $k_{2,\pm}$ compared to $k_{1,\pm}$
\cite{kodama2004young}.
Further, 
each of the above three cases corresponds to a different kind of soliton interaction \cite{biondini2007line}. 
Specifically, and taking $k_{1,-}<k_{2,-}$ without loss of generality, 
there are three different classes of solutions,
corresponding to the following three possible cases:
\vspace*{-1ex}
\bse
\label{e:KPpairwiseinequalities}
\begin{enumerate}
\item[(i)]
Ordinary soliton interaction:  
$k_{1,+} < k_{2,-}$ (corresponding to no overlap between $k_{1,\pm}$ and $k_{2,\pm}$), 
which occurs if and only if 
\be \label{ord}
(c_2 - c_1)/(2\sqrt3) + a_1+ a_2 < 0 \,.
\ee
This corresponds to ``standard'' soliton interactions: the two solitons emerge unscathed after their interaction, except for a positive phase shift.
\item[(ii)] 
Resonant soliton interaction: 
$k_{2,-} < k_{1,+} < k_{2,+}$
(corresponding to partial overlap between $k_{1,\pm}$ and $k_{2,\pm}$), 
which occurs if and only if 
\be
-(a_1+a_2) < (c_2 - c_1)/(2\sqrt3) < a_2- a_1.
\ee
Resonant two-soliton interactions produce a complex, time-dependent 
box-shaped pattern \cite{biondini2003family}, in which, 
at each vertex, two solitons merge into one, producing a local $Y$ shape characterized by 
a so-called Miles soliton resonance \cite{miles1977diffraction,medina2002n}.
\item[(iii)] 
Asymmetric soliton interaction: 
$k_{1,+}> k_{2,+}$ 
(corresponding to ``total overlap'' between $k_{1,\pm}$ and $k_{2,\pm}$), 
which occurs if and only if 
\be\label{assym}
c_2 - c_1 > 2\sqrt3\,(a_1 + a_2)\,.
\ee
Asymmetric interactions are similar to ordinary ones, the only difference being that solitons incur a negative phase shift.
\end{enumerate}
\ese 

The inequalities~\eqref{e:KPpairwiseinequalities} 
cover the whole soliton parameter space $(a_1,a_2,c_1,c_2)$ for the KPII equation.
On the other hand, the above trichotomy has important consequences in terms of the reduction to the Boussinesq equation,
since only ordinary and asymmetric pairwise 2-soliton interactions lead to stationary solutions.
One must therefore impose a pairwise non-resonance condition on the parameters of 
the individual solitons. 
That is, one must impose that each pair of phase parameters has either 
no overlap or total overlap with every other pair. 
Moreover, when the relation \eqref{e:Bamplitude} is taken into account,
\eqref{e:KPpairwiseinequalities} imposes additional restrictions on the allowable choices of slopes,
with the result that not all possible pairs $c_1$ and $c_2$ are admissible.
This phenomenon was first pointed out in \cite{lambert1987soliton} and put in the context of the KP formalism in
\cite{bogdanov2002boussinesq},
where an alternative formulation of the multi-soliton solutions of the Boussinesq equation was given,
and it was shown that $N$-soliton solutions of Eq.~\eqref{e:Boussinesq} can be expressed via Eq.~\eqref{e:utau},
where now the tau function is the determinant of an $N\times N$ matrix, namely,
\be\label{e:tauB}
\tau(x,y,t=0) = \det (I + M)\,, 
\ee
where $I$ is the $N\times N$ identity matrix and $M = (M_{jk})$, with
\vspace*{-1ex}
\begin{equation}\label{e:BouM}
    M_{jk} = \frac{C_j}{\mu_j - \lambda_k}\exp\left[\frac{\mu_j-\lambda_j}{2}\bigg(x - \frac{\sqrt 3}{2}(\mu_j+\lambda_j)y\bigg)\right],
\end{equation}
and where the real-valued parameters $\mu_j$ and $\lambda_j$ are related by the constraint
\vspace*{-1ex}
\begin{equation}\label{e:mulambda}
    2\mu_j + \lambda_j \pm \sqrt{4-3\lambda_j^2} = 0\,.
\end{equation}
By computing the $1$-soliton solution explicitly using the above formalism, 
one recovers Eq.~\eqref{e:KPsoliton} with $a = (\mu-\lambda)/4$ and
$c = {\sqrt 3}(\mu+\lambda)/2$,
which shows that the parameters $\mu$ and $\lambda$ are simply rescaled phase parameters of the KPII equation:
$\lambda = 2k_-$ and $\mu = 2k_+$.
One can then check that this yields the correct relation between amplitude and velocity, namely Eq.~\eqref{e:Bamplitude},
which is equivalent to Eq.~\eqref{e:mulambda}.

\medskip
\subsection{Interactions in two-component gas}
For the two-component (bi-chromatic) SG we generally have $M^{\rm l} + M^{\rm r}=2$ so the substitution of the delta-function ansatz \eqref{polychrom} in the kinetic equation \eqref{e:RarContEq} yields the system of hydrodynamic conservation laws \eqref{e:RarContEq_Ht}, which we reproduce here:
\begin{equation}\label{cons_gen}
    (F^{\alpha}_1)_y +  (s^{\alpha}_1F^{\alpha}_1)_x = 0, \ \  (F^{\beta}_2)_y +  (s^{\beta}_2 F^{\beta}_2)_x = 0
    \end{equation}
complemented by the algebraic equations of state
 \bse \label{s_12}
 \begin{gather}
s^{\alpha}_1= c^{\alpha}_{1} + \frac{ \big[c^{\alpha}_{1}-c^{\beta}_{2}\big]\Delta_{12} F^{\beta}_2}{1-\big[\Delta_{12} F^{\beta}_2+ \Delta_{21} F^{\alpha}_1\big]},\\
\quad s^{\alpha}_{2}= c^{\beta}_{2}-\frac{ \big[c^{\alpha}_{1} -c^{\beta}_{2} \big]\Delta_{21} F^{\alpha}_1}{1-\big[\Delta_{12} F^{\beta}_2+ \Delta_{21} F^{\alpha}_1\big]}. 
\end{gather}
\ese
 It is important to stress that the bichromatic reductions are subject to the constraint  $1-(\Delta_{12} F^{\beta}_2+ \Delta_{21} F^{\alpha}_1)>0$ which implies that the denominators in \eqref{s_12} must be positive \cite{congy2024riemann}.

\subsection{Refraction slopes}
To extract the slope of a single line r-soliton refracted by a SG (either r- or l-) we set $\Fr_1=0$ in Eq. \eqref{s_12}a. As a result, the co-directional r-r refraction slope parameter is given by (see Eq.~\eqref{codir_ref} in the main text)
\begin{equation}\label{codir_ref_sup}
\ceffr_1 =\crr_{1}+\frac{\big[\crr_{1}- \crr_{2}\big]\Delta^{\rm co}(\ar_1,\ar_2)\Fr_{2}}{1-\Delta^{\rm co}(\ar_1,\ar_2)\Fr_{2}}
\end{equation}
while for the counter-directional r-l refraction we obtain (see Eq.~\eqref{ctdir_ref} in the main text)
\begin{equation}\label{ctdir_ref_sup}
\ceffr_1 =\crr_{1}+ \frac{\big[\crr_{1}- \cl_{2}\big]\Delta^{\rm ct}(\ar_1,\al_2)\Fl_{2}}{1-\Delta^{\rm ct}(\ar_1,\al_2)\Fl_{2}}     
\end{equation}

\subsection{Derivation of the slopes of contact discontinuities}
The Riemann problem describing the oblique interference of two monochromatic SGs  is specified by the step conditions at $y=0$ for the system \eqref{cons_gen}
\begin{align} \label{riem_cond1}
		F^{\alpha}_1(x,0)=F^{\alpha}_{10}, \ F^{\beta}_{2}(x,0)=0 \ \ \text{if } x<0,\\[2mm]
		F^{\alpha}_1(x,0)=0, \  F^{\beta}_2(x,0)=F^{\beta}_{20} \ \ \text{if } x> 0.
\end{align}
The dependence of the ``initial'' velocities $s^{\alpha}_{1}(x,0)$, $s^{\beta}_{2}(x,0)$ on the densities $F^{\alpha}_1(x,0), F^{\beta}_2(x,0)$  is defined by the relations \eqref{s_12}.
Due to the scale invariance of the Riemann problem, the solution is a self-similar distribution $F^{\alpha}_{1}(x/y)$, $F^{\beta}_{2}(x/y)$. Because of linear degeneracy of the hydrodynamic 
system \eqref{cons_gen}, \eqref{s_12} the only admissible similarity solutions are constant states
separated by propagating contact discontinuities, see for
instance \cite{rozhdestvenskii_systems_1983}. Discontinuous weak
solutions are physically admissible here since the kinetic equation
describes the conservation of the number of solitons within any given spectral interval, and the Rankine-Hugoniot type conditions can be imposed
to ensure the conservation of the number of solitons across the
discontinuities. As a result, the solution of the Riemann problem \eqref{cons_gen}, \eqref{init_r} for each component is composed of
$3$ constant states separated by $2$ contact discontinuities \cite{el2005kinetic, carbone_macroscopic_2016, el2021soliton}:
\vspace*{-0.6ex}
\begin{equation}\label{init_r}
	(F^{\alpha}_1(x,y), F^{\beta}_2(x,y)) =
	\begin{cases}
		 (F^{\alpha}_{10}, 0), &x/y<c^-,\\[2mm]
		(F^{\alpha}_{1c}, F^{\beta}_{2c}),& c^- < x/y < c^+,\\[2mm]
		(0, F^{\beta}_{20}), & x/y> c^+; 
	\end{cases}
 \end{equation}
where the component densities $F^{\alpha}_{1c}$ and $F^{\beta}_{2c}$ in the interference region $c^-y <x <c^+y$ and the inverse slopes of the contact discontinuities $c^\pm$ are found for the Rankine-Hugoniot conditions $-c \llbracket F^{\gamma}_i \rrbracket + \llbracket F^{\gamma}_i s^{\gamma}_i \rrbracket=0$, $i=1,2$ applied across $x=c y$. Here $c$ is either $c^-$ (left discontinuity) or $c^+$ (right discontinuity) and $\llbracket F \rrbracket$ denotes the jump of $F$ across the discontinuity.
As the result one obtains 
\bse \label{contactSM}
\begin{align}
&c^+=c^{\alpha}_{1}+\frac{\big[c^{\alpha}_{1}-c^{\beta}_{2}\big]\Delta_{1,2}F^{\beta}_{20}}{1-\Delta_{2,1}F^{\beta}_{20}}, \\
&c^-=c^{\beta}_{2}-\frac{\big[c^{\alpha}_{1}-c^{\beta}_{2}\big]\Delta_{2,1}F^{\alpha}_{10}}{1-\Delta_{1,2}F^{\alpha}_{10}}.
\end{align}
\ese
The component densities in the interference region  are  given by
\bse\label{plateauSM}
\begin{align}
&F^{\alpha}_{1c}=\frac{F^{\alpha}_{10}\big[1-\Delta_{1,2}F^{\beta}_{20}\big]}{1-\Delta_{1,2}\Delta_{2,1}F^{\alpha}_{10}F^{\beta}_{20}}, \\ 
&F^{\beta}_{2c}=\frac{F^{\beta}_{20}\big[1-\Delta_{2,1}F^{\alpha}_{10}\big]}{1-\Delta_{1,2}\Delta_{2,1}F^{\alpha}_{10}F^{\beta}_{20}}.
\end{align}
\ese
See the main text for the interpretation of these formulae in the context of 2D oblique interference of SGs.

\medskip
\subsection{Correlations}
We now discuss how to obtain correlations in 2D stationary KPII gases, using standard GHD results \cite{doyon2018exact} first applied to Boussinesq SGs in \cite{bonnemain_soliton_2024}. We refer the interested reader to these two references for a more in depths discussion as, here, we will only present the bare minimum required to obtain results from the main text.

We first need to define a quantity that is fundamental in GHD: the occupation function. 
Like the DOSs, it comes in two flavours, the \emph{left-} and \emph{right-occupation}, $n^{\rm l}$ and $n^{\rm r}$, defined by
\begin{equation}
    \frac{2\pi f^\alpha(a)}{n^\alpha(a)} = a +  \sum_{\gamma = \alpha,\beta} \int_{A^\gamma} \dd a' f^\gamma(a') \Delta(a,a') \ ,  
\end{equation}
where $\alpha$ is either l or r, and $\beta$ is r or l respectively. This quantity is used to define the left- and right-dressing operations
\begin{equation}
    g^{\alpha, {\rm dr}}(a) = g(a) + \sum_{\gamma = \alpha,\beta} \int_{A^\gamma} \frac{\dd a'}{2\pi}\Delta(a,a')n^\gamma(a')g^{\gamma, {\rm dr}}(a') \; ,
\end{equation}
for any smooth function $g(a)$. The horizontal covariance of the wave-field is then given as
\vspace*{-1ex}
\begin{multline}
    \int_{\mathbb R} \dd x \langle u(x,y)u(0,y) \rangle^c \\
    = \sum_{\gamma = \alpha,\beta} \int_{A^\gamma} \dd a f^\gamma(a;0,y)\left(h_0^{\gamma, {\rm dr}}(a;0,y)\right)^2,
\end{multline}
where $h_0(a)=4a$. This particularly simple form comes from the fact that the wave-field $u$ is also the density of the first conserved quantity associated with the Boussinesq equation $Q_0 = \int \dd x \,u(x,y)$, $\partial_y Q_0 = 0$, and $h_0(a)\equiv 4a$ corresponds to $Q_0$ evaluated for a single-soliton solution of amplitude $a$ 
(cf.\ \cite{bonnemain_soliton_2024} for details). 
One can also evaluate two-point correlations as
\vspace*{-1ex}
\begin{multline}
    %\frac{1}{2L}\int_{-L}^L \dd x &\langle u(x+y\xi ,y)u(0,0) \rangle^c \\
    \frac{1}{2L}\int_{-L}^L \dd x' \langle u(x+x',y)u(0,0) \rangle^c 
    \\
    =\sum_{\gamma = \alpha,\beta} \int_{A^\gamma}\dd a \, \delta\left(x-s^\gamma(a;0,0) y\right)f^\gamma(a;x,y) \times 
    \\
    h_0^{\gamma,\rm dr}(a;x,y)h_0^{\gamma, \rm dr}(a;0,0)\\
    \sim \frac{1}{y}\sum_{\gamma = \alpha,\beta}\sum_{a\in a^{\gamma}_*} \frac{f^\gamma(a;0,0)}{|{\partial_a s^\gamma}(a;0,0)|}\big(h_0^{\gamma, {\rm dr}}(a;0,0)\big)^2,
\end{multline} 
for $y\gg L\gg \ell$, where $\ell = \left[\sum_{\gamma = \alpha,\beta} \int_{A^\gamma}\dd a f^\gamma(a;0,0) \right]^{-1}$ quantifies the average inter-soliton spacing along the $x$~direction, and $a^{\gamma}_* = \{a: s^\gamma(a;0,0) = x/y\}$. 

\begin{figure*}[t!]
\centerline{\includegraphics[width=1.\textwidth]{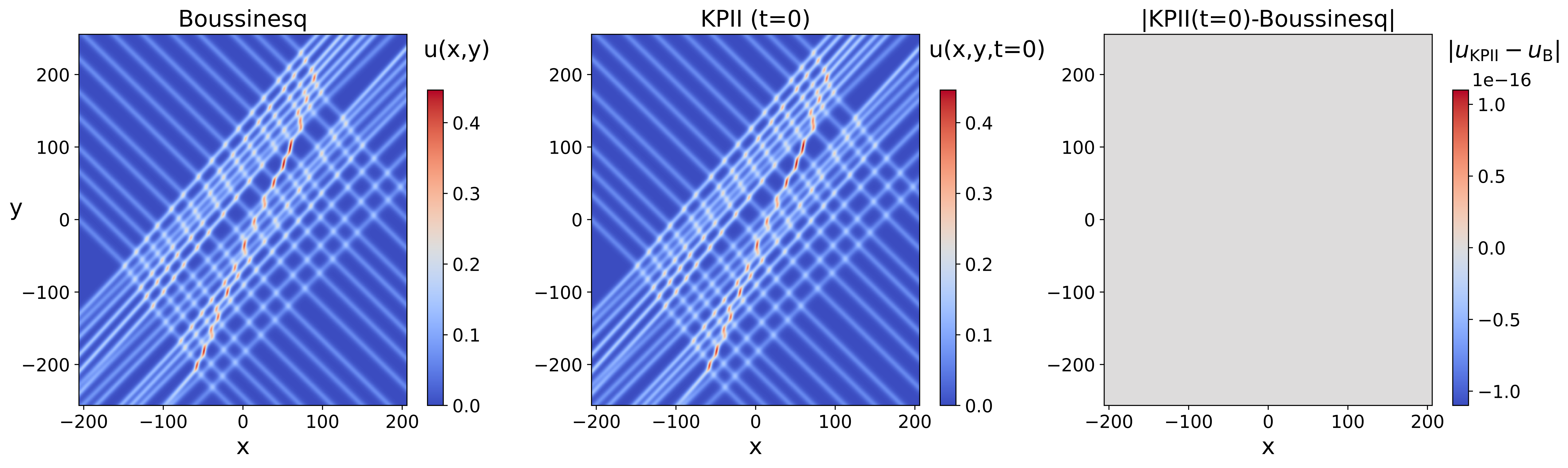}}
\medskip
\centerline{\includegraphics[width=1.\textwidth]{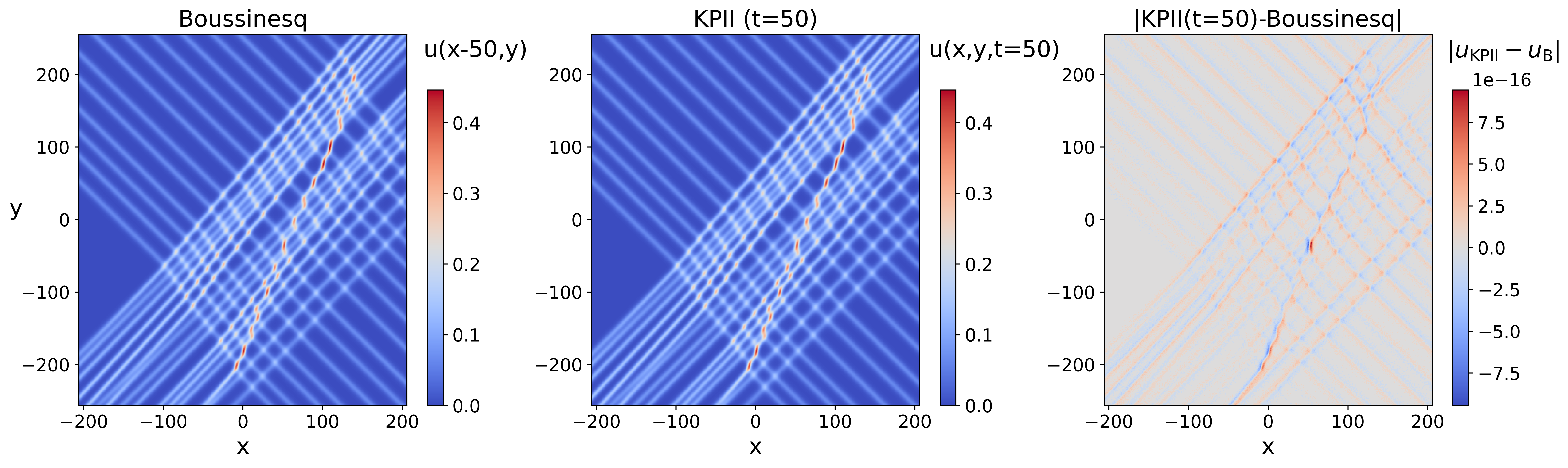}}
 \caption{Comparison of 2D line SG interference implemented via exact $N$-soliton solution of the Boussinesq (left) and KPII (centre) equations and the absolute value of the difference between the solutions generated using the two different algorithms (right).  The positively sloped gas is composed of $N^{\rm r}=15$ solitons with parameter $a^{\rm r}$ uniformly distributed in the range $a^{\rm r}\in [0.175, 0.281]$. The negatively sloped gas is composed of $N^{\rm l}=15$ solitons with parameter $a^{\rm l}$ uniformly distributed in the range $a^{\rm l}\in [0.175, 0.183]$. Top row: KPII solution generated at $t=0$ compared with the Boussinesq solution.  Note that the difference is identical to zero up to the accuracy used to save the data.
 Bottom row: KPII solution generated at $t=50$ compared with the Boussinesq solution shifted by $\Delta x = 50$, according to the Galilean-boost adopted. Note that the difference is in the order of magnitude of the accuracy used to save the data.}
 \label{fig:KPvsB}
 \end{figure*}

Putting all of this in the context of the r-r and r-l bi-chromatic reductions of the main text yields for the occupation function
\begin{subequations}
\begin{align}
    &n^{\rm r}(a) = 2\pi\frac{F_1^{\rm r}\delta(a-a_1^{\rm r})+F_2^{\rm r}\delta(a-a_2^{\rm r})}{a+ F_1^{\rm r}\Delta^{\rm co}(a,a_1^{\rm r})+F_2^{\rm r}\Delta^{\rm co}(a,a_2^{\rm r})} \ , \\
    &n^{\rm l}(a) = 0 \ ,
\end{align} 
\end{subequations}
in the r-r case, and
\begin{subequations}
\begin{align}
    &n^{\rm r}(a) = 2\pi\frac{F_1^{\rm r}\delta(a-a_1^{\rm r})}{a+ F_1^{\rm r}\Delta^{\rm co}(a,a_1^{\rm r})+F_2^{\rm l}\Delta^{\rm ct}(a,a_2^{\rm l})} \ , \\ &n^{\rm l}(a) = 2\pi\frac{F_2^{\rm l}\delta(a-a_2^{\rm l})}{a+ F_1^{\rm r}\Delta^{\rm ct}(a,a_1^{\rm r})+F_2^{\rm l}\Delta^{\rm co}(a,a_2^{\rm l})} \ ,
\end{align}
\end{subequations}
in the r-l one. Moreover, since
\begin{equation}
    |\Delta(a,a')| \to \infty \qquad \text{as} \qquad {a\to a'}  \; , 
\end{equation}
for any admissible amplitude $a'$, we have, e.g.
\begin{equation}
\lim_{a\to a'}   \frac{F_1^{\rm r}\Delta^{\rm co}(a,a')}{a+ F_1^{\rm r}\Delta^{\rm co}(a,a')+F_2^{\rm r}\Delta^{\rm co}(a,a')} = 1 \; ,
\end{equation}
hence we may evaluate the dressing of $h_0(a)$. First, in the r-r case, we have
\begin{equation}
    h_0^{\rm r, dr}(a) = h_0(a) + h_0^{\rm r, dr}(a_1^{\rm r}) + h_0^{\rm r, dr}(a_2^{\rm r}) \; , 
\end{equation}
and by fixing either $a=a_1^{\rm r}$ and $a=a_2^{\rm r}$ we eventually find
\begin{subequations}
\begin{align}
    &h_0^{\rm r, dr}(a_1^{\rm r}) = -h_0(a_2^{\rm r}) = -4a_2^{\rm r}  \ , \\
    &h_0^{\rm r, dr}(a_2^{\rm r}) = -h_0(a_1^{\rm r})= -4a_1^{\rm r} \ .
\end{align}
\end{subequations}
Similarly, we have, in the r-l case
\begin{subequations}
\begin{align}
    &h_0^{\rm r, dr}(a_1^{\rm r}) = -h_0(a_2^{\rm l})=-4a_2^{\rm l} \ , \\
    &h_0^{\rm l, dr}(a_2^{\rm l}) = -h_0(a_1^{\rm r})=-4a_1^{\rm r} \ .
\end{align}
\end{subequations}
Ultimately, using these results, we can compute the horizontal covariance and the two-point correlation function, in order to recover Eqs.~\eqref{bicov} of the main text.

Note that in the case of two-point correlations, the fact that the generated gas is not strictly polychromatic matters; indeed, instead of the delta-functional ansatz \eqref{polychrom} we have
\begin{equation}
    \tilde f^\alpha(a;x,y) = \sum^{M^\alpha}_{i=1} \frac{F^\alpha_i(x,y)}{\epsilon_g}\Theta(|a_i^\alpha-a|) \ ,
\end{equation}
where $\Theta(a)$ is the Heaviside function, such that $\int_{A^\alpha} \dd a f^\alpha(a;x,y) = \int_{A^\alpha} \dd a \tilde f^\alpha(a;x,y)$, $\forall \ (x,y)$. This becomes significant because two-point correlations have to be evaluated over sufficiently large scales, on which the gas cannot be considered homogeneous anymore. Over such large scales, the slight difference between the spectral parameter of a given soliton within the gas and the reference value $a_i^\alpha$ may translate, through the multiple scattering shifts it incurs, into a meaningful deviation from its reference trajectory $s^\gamma(a_i^\alpha) = x/y$. As such, we can define a ``light cone" by computing the effective velocities $s^\gamma_{i, \ -}$ and $s^\gamma_{i, \ +}$ a soliton of amplitude $a_{i, \ -}^{\alpha}=a_i^\alpha- \epsilon_g/2$ and one of amplitude $a_{i, \ +}^{\alpha}=a_i^\alpha+ \epsilon_g/2$ would have if they interacted with the polychromatic gas
\begin{equation}
\begin{aligned}
    &s^{\gamma}_{i, \ -} = \frac{c(a_{i, \ -}^{\gamma}) - s^\alpha_1 F^{\alpha}_{1c}\Delta(a_{i, \ -}^{\gamma},a_1^\alpha) - s^\beta_2F^\beta_{2c}\Delta(a_{i, \ -}^{\gamma},a_2^\beta)}{1-F^{\alpha}_{1c}\Delta(a_{i, \ -}^{\gamma},a_1^\alpha) - F^\beta_{2c}\Delta(a_{i, \ -}^{\gamma},a_2^\beta)} \ ,\\
    &s^\gamma_{i,\ +} = \frac{c(a_{i,\ +}^{\gamma}) - s^\alpha_1 F^{\alpha}_{1c}\Delta(a_{i,\ +}^{\gamma},a_1^\alpha) - s^\beta_2F^\beta_{2c}\Delta(a_{i,\ +}^{\gamma},a_2^\beta)}{1-F^{\alpha}_{1c}\Delta(a_{i,\ +}^{\gamma},a_1^\alpha) - F^\beta_{2c}\Delta(a_{i,\ +}^{\gamma},a_2^\beta)} \ ,
\end{aligned}
\end{equation}

where we assumed that modifying the amplitude of a single soliton from the gas by adding or subtracting $\epsilon_g/2$ does not significantly affect the weights of the two components of the bi-chromatic gas defined by Eqs~\eqref{plateauSM}. The velocity of any soliton from the gas should verify $s^\gamma_{i,\ -} < s^\gamma(a;x,y) < s^\gamma_{i,\ +}$, which means that, given a known position $x$ at which a soliton intersects with the horizontal at $y=0$, the position at which it will intersect the horizontal at $y=Y$ is known with some uncertainty $\delta x(Y,\epsilon_g) = |s^\gamma_{i,\ +}-s^\gamma_{i,\ -}|Y$. In the main text, when computing the two-point correlations \eqref{bitwopoint}, $\epsilon_g$ must be chosen to ensure that $\delta x(Y,\epsilon_g) \ll Y$.

\subsection{Numerical generation of stationary multisoliton KPII solutions}
The numerical generation of $N$-soliton solutions adapted to approximate the stationary SG, for $N\gg 1$, relies on the Wronskian formulation \eqref{e:utau}  where the function $\tau(x,y,t)$ is either given by \eqref{e:tauB}, for the Boussinesq equation, or \eqref{e:tauNM}, for the KPII equation. In the latter case, it is more advantageous to consider the formulation $\tau = \det(K\,\e^{\Theta}G^T)$ than to adopt the Binet-Cauchy formula, that relies on the Van der Monde determinants (\ref{e:tauNM}), as the number of non-zero minors increases exponentially with the number of solitons, making the algorithm unsuitable for the numerical study of solutions with $N\gg1$. 
Note that, when it comes to the Boussinesq equation, 
in Eq.~\eqref{e:BouM} we have $C_j=\left(\mu_j-\lambda_j\right)\exp\big[-\frac12(\mu_j-\lambda_j)x_{0j}\big]$, 
where $x_{0j}$ represents the position of the maximum of the line soliton solution at $y=0$.

As previously discussed, stationary solutions of the Galilean-boosted KPII equation can be identified with solutions of the Boussinesq equation. To this end,  the map between the two expressions of $\tau(x,y,t=0)$, for the Boussinesq and KPII equations, is detailed below.
The matrix $G$ in \eqref{e:tauNM} can generically be put in the form
\begin{equation}
G=\begin{pmatrix}
g_{1,1} & g_{1,2} & 0 & 0 & \cdots & 0 & 0 \\  0 & 0 & g_{2,3} & g_{2,4} & \cdots & 0 & 0 \\ \vdots & \vdots & \vdots  &  \vdots & \ddots &  \vdots &  \vdots \\ 0 &0 & \cdots & \cdots & 0  &g_{N,2N-1} & g_{N,2N}
\end{pmatrix} \; ,
\end{equation}
and, comparing  \eqref{e:tauNM} with  \eqref{e:BouM}, we obtain the map:
\begin{equation}
g_{n,2n} = -c_n \frac{g_{n,2n-1}}{\mu_n-\lambda_n}\prod_{n\neq m}\frac{\lambda_n-\lambda_m}{\mu_n-\lambda_m}
\end{equation}
where $c_n=\lambda_n-\mu_n $. Without loss of generality, we can assume $g_{n,2n-1}=1$ for $n=1,2\ldots N-1$ and the element $g_{N,2N-1}$ is given by:
\vspace*{-1ex}
\begin{equation}
g_{N,2N-1}=\prod_{n=N}^{1}\prod_{m=1}^{n-1} \sgn(\lambda_n-\lambda_m).
\end{equation}
Note that, once the map is established, the solutions generated by the two 
 different algorithms are identical, see  Fig.~\ref{fig:KPvsB}.

As is the case with the schemes for the numerical construction of the $N$-soliton solution for NLS and KdV using the Darboux transformation  \cite{gelash2018strongly,congy2023dispersive}, this algorithm is subject to round-off errors during summation of exponentially small and large values.  Following the procedure developed in  \cite{gelash2018strongly}, high--precision arithmetic routines have been implemented to overcome the numerical accuracy problems.

\subsection{Numerical realization of monochromatic SGs}
In practice, the monochromatic SGs discussed in the main text are approximated by an $N$-soliton solutions consisting of a  “cluster” of $N$ solitons with parameters $a_{i}$ uniformly distributed in a small region of width $\delta_g$ centred around $a_g$. Additionally, the randomness of the gas is achieved by uniformly distributing $x_{0j}$ on some interval of width $\delta_x$. The width $\delta_x$ non-trivially determines the density $F_g$ of the gas. This is computed numerically as the ratio between the selected number of solitons $N$ in the gas over its spatial extension $L$ in the $x$ direction: $F_g = N/L$.

%%%%%%%%%%%%%%%%%%%%%%%%%%%%%%%%%%%%%%%%%%%%%%%%%%%%%%%%%%%%%%%%%%%%%%%%%%%%%%%%%%%%%%%%%%%%%%%%%%%
%%%%%%%%%%%%%%%%%%%%%%%%%%%%%%%%%%%%%%%%%%%%%%%%%%%%%%%%%%%%%%%%%%%%%%%%%%%%%%%%%%%%%%%%%%%%%%%%%%%
%%%%%%%%%%%%%%%%%%%%%%%%%%%%%%%%%%%%%%%%%%%%%%%%%%%%%%%%%%%%%%%%%%%%%%%%%%%%%%%%%%%%%%%%%%%%%%%%%%%
\bibliographystyle{apsrev4-1}
\bibliography{Biblio,Biblio-2019-01-03,Bibliography}
\end{document}